\documentclass[sigconf]{acmart}

\usepackage{makecell}
\usepackage{arydshln}
\usepackage{multirow}
\usepackage{bm}
\usepackage{threeparttable}

\usepackage[table]{xcolor}

\AtBeginDocument{%
  }

\setcopyright{acmlicensed}
\copyrightyear{2026}
\acmYear{2026}
\acmDOI{XXXXXXX.XXXXXXX}
\acmConference[Conference acronym 'XX]{Make sure to enter the correct
  conference title from your rights confirmation email}{June 03--05,
  2026}{Woodstock, NY}
\acmISBN{978-1-4503-XXXX-X/2018/06}

\begin{document}

\title{MECO: A Multimodal Dataset for Emotion and Cognitive Understanding in Older Adults}


\author{Hongbin Chen}
\affiliation{%
  \institution{Nanjing Medical University}
  \city{Nanjing}
  \country{China}
}
\email{hongbinchen@stu.njmu.edu.cn}

\author{Jie Li}
\affiliation{%
  \institution{ Nanjing Medical University}
  \city{Nanjing}
  \country{China}
}
\email{jerry@njmu.edu.cn}

\author{Wei Wang}
\affiliation{%
  \institution{Nanjing Medical University}
  \city{Nanjing}
  \country{China}
}
\email{bmeww@njmu.edu.cn}

\author{Siyang Song}
\affiliation{%
  \institution{University of Exeter}
  \city{Exeter}
  \country{U.K.}
}
\email{s.song@exeter.ac.uk}

\author{Xiao Gu}
\affiliation{%
  \institution{University of Oxford}
  \city{Oxford}
  \country{U.K.}
}
\email{xiao.gu@eng.ox.ac.uk}

\author{Jianqing Li}
\authornote{Corresponding author.}
\affiliation{%
  \institution{Nanjing Medical University}
  \city{Nanjing}
  \country{China}
}
\email{jqli@njmu.edu.cn}

\author{Wentao Xiang}
\authornotemark[1]
\affiliation{%
  \institution{Nanjing Medical University}
  \city{Nanjing}
  \country{China}
}
\email{xiangbmu@njmu.edu.cn}

\renewcommand{\shortauthors}{Chen et al.}

\begin{abstract}
    While affective computing has advanced considerably, multimodal emotion prediction in aging populations remains underexplored, largely due to the scarcity of dedicated datasets. Existing multimodal benchmarks predominantly target young, cognitively healthy subjects, neglecting the influence of cognitive decline on emotional expression and physiological responses. To bridge this gap, we present MECO, a \textbf{M}ultimodal dataset for \textbf{E}motion and \textbf{C}ognitive understanding in \textbf{O}lder adults. MECO includes 42 participants and provides approximately 38 hours of multimodal signals, yielding 30,592 synchronized samples. To maximize ecological validity, data collection followed standardized protocols within community-based settings. The modalities cover video, audio, electroencephalography (EEG), and electrocardiography (ECG). In addition, the dataset offers comprehensive annotations of emotional and cognitive states, including self-assessed valence, arousal, six basic emotions, and Mini-Mental State Examination cognitive scores. We further establish baseline benchmarks for both emotion and cognitive prediction. MECO serves as a foundational resource for multimodal modeling of affect and cognition in aging populations, facilitating downstream applications such as personalized emotion recognition and early detection of mild cognitive impairment (MCI) in real-world settings. The complete dataset and supplementary materials are available at \url{https://maitrechen.github.io/meco-page/}.
\end{abstract}



\begin{teaserfigure}
\centering
\includegraphics[width=\linewidth]{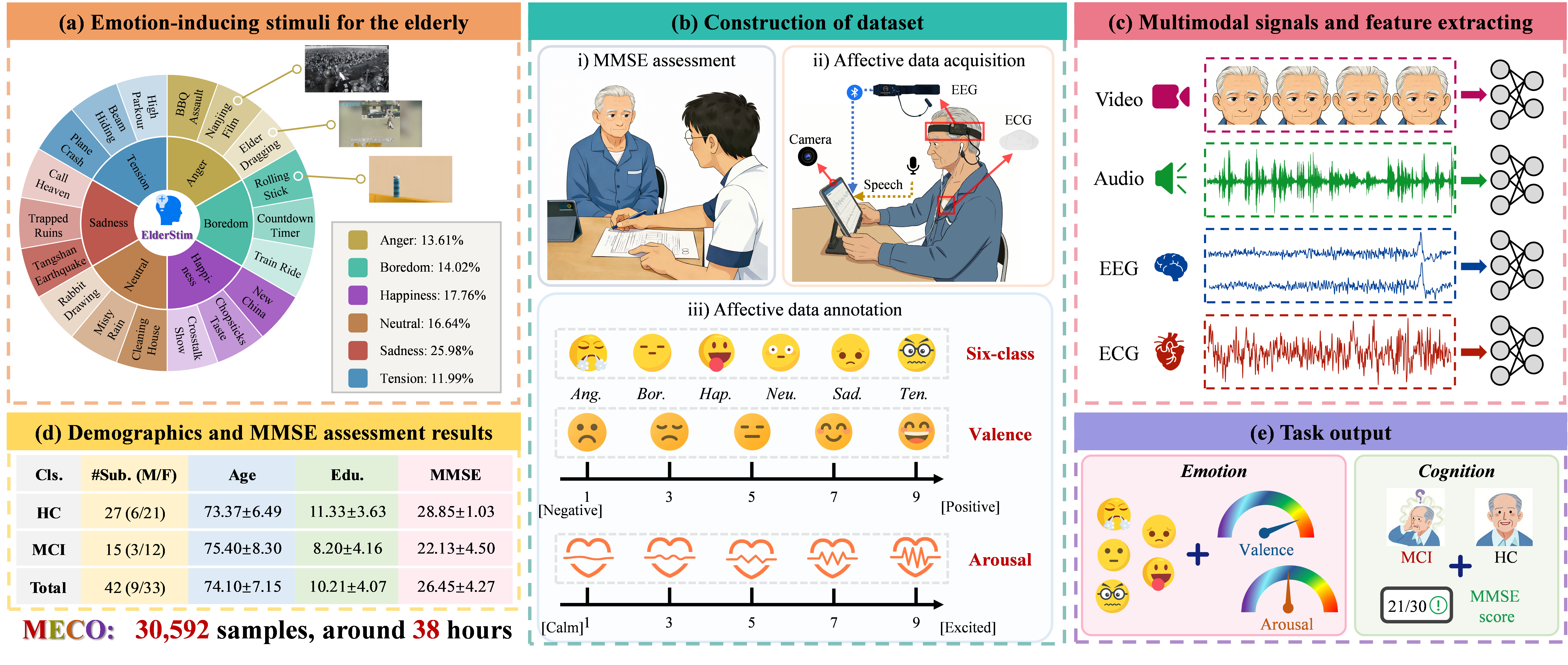}
\caption{Overview of MECO dataset. (a) Emotion-inducing video stimuli for older adults. (b) Data acquisition protocol, encompassing cognitive assessment, synchronized multimodal recording, and self-assessed annotations. (c) Extracted multimodal signals and corresponding feature representations. (d) Downstream tasks for emotion and cognitive prediction. (e) Demographic characteristics of the study subjects, highlighting group-specific differences.}
\Description{A four-part overview of the MECO dataset. The first part shows emotion-eliciting video stimuli presented to older adults. The second part illustrates the data collection workflow, including Mini-Mental State Examination cognitive assessment, synchronized acquisition of multimodal signals using wearable sensors, and self-reported emotional annotations. The third part presents the recorded modalities and corresponding feature extraction processes across behavioral and physiological signals. The fourth part depicts downstream tasks, including classification and regression for both emotional and cognitive states.}
\label{fig-overview}
\end{teaserfigure}

\received{20 February 2007}
\received[revised]{12 March 2009}
\received[accepted]{5 June 2009}

\maketitle

\section{Introduction}
With the rapid growth of the global aging population, understanding affective states in older adults is increasingly critical for mental health monitoring and cognitive assessment \cite{Beard2016}. Cognitive decline, such as mild cognitive impairment (MCI), can markedly alter emotional expression and physiological responses \cite{Ismail2015}, posing significant challenges for reliable affective analysis in older adults. Automated and quantitative modeling of these coupled factors holds substantial potential for improving early detection and intervention in geriatric care \cite{John2018}.

Despite recent advances in multimodal affective computing, the development of robust systems for older adults remains constrained by intersecting gaps in existing datasets. Most public benchmarks are dominated by young, cognitively intact individuals and assume modality congruence, where outward facial expressions synchronously reflect internal arousal \cite{Mauss2005,Katsigiannis2018,WeiLongZheng2015,Koelstra2012}. This assumption fails to capture atypical affective manifestations in older populations, particularly those with MCI. Patients with MCI frequently exhibit facial apathy \cite{Robert2009}, leading to a disconnect between blunted outward expressions and active internal physiological arousal. This renders traditional visually-driven datasets inadequate and causes existing models to misinterpret emotional states. Moreover, the intrinsic interplay between cognition and emotion is largely overlooked, as current datasets typically annotate these states in isolation \cite{Luz2020,Soleymani2012,Park2020,Lee2024}, ignoring the clinical reality that cognitive decline actively modulates emotional reactivity, making it difficult to investigate how progressive cognitive degradation reshapes multimodal emotional representations. To capture these complex interactions, models require synchronized behavioral (video) and physiological modalities-electroencephalography (EEG) and electrocardiography (ECG)-alongside cognitive assessments \cite{BagherZadeh2018,Jiang2020,Yang2025}.

\begin{table*}[!t]
\footnotesize
\centering
\caption{Review of representative multimodal emotion datasets. "N/A" denotes information not available.}

\begin{tabular}{cccccccc}
\toprule
\textbf{Dataset} & \textbf{Age (Avg.)} & \textbf{{\#}Subjects (M/F)} &
\textbf{Length} & 
\textbf{Label} & 
\textbf{Primary Modality} & 
\textbf{Language} & \textbf{Source} \\
\midrule

IEMOCAP \cite{Busso2008}
    & N/A
    & 10 (5/5)
    & 12 h
    & Emotion
    & Audio, Video
    & English
    & In the lab \\

DFEW \cite{Jiang2020}
    & N/A
    & N/A
    & N/A
    & Emotion
    & Audio, Video
    & N/A
    & In the wild \\

DREAMER \cite{Katsigiannis2018}
    & 22--33 (26.6)
    & 23 (14/9)
    & 7 h
    & Emotion
    & EEG, ECG
    & N/A
    & In the lab \\

SEED-IV \cite{WeiLongZheng2015}
    & 18--30 (23.3)
    & 15 (7/8)
    & 30 h
    & Emotion
    & EEG, EOG
    & Chinese
    & In the lab \\

DEAP \cite{Koelstra2012}
    & 19--37 (26.9)
    & 32 (16/16)
    & 21 h
    & Emotion
    & Video, EEG
    & N/A
    & In the lab \\

MAHNOB-HCI \cite{Soleymani2012}
    & 19--40 (26.1)
    & 27 (11/16)
    & 12 h
    & Emotion
    & Audio, Video, EEG
    & English
    & In the lab \\

ElderReact \cite{Ma20191}
    & N/A
    & 46 (20/26)
    & 2 h
    & Emotion
    & Audio, Video
    & English
    & In the wild \\

\midrule

EMOTyDA \cite{Saha2020}
    & N/A
    & N/A
    & 22 h
    & Emotion, Intention
    & Audio, Video, Text
    & English
    & TV+In the lab \\

MINE \cite{Yang2025}
    & N/A
    & N/A
    & 22 h
    & Emotion, Intention
    & Audio, Video, Image, Text
    & English
    & In the wild \\

ASCERTAIN \cite{Subramanian2018}
    & N/A (30.0)
    & 58 (37/21)
    & 46 h
    & Emotion, Personality
    & Video, EEG, ECG, GSR
    & English
    & In the lab \\
    
AMIGOS \cite{MirandaCorrea2021}
    & 21--40 (28.3)
    & 40 (27/13)
    & 69 h
    & Emotion, Personality, Mood
    & Video, EEG, ECG, GSR
    & English
    & In the lab \\    
\midrule

\textbf{MECO (Ours)}
    & \textbf{57--85} (\textbf{74.1})
    & \textbf{42 (9/33)}
    & \textbf{38 h}
    & \textbf{Emotion, Cognition}
    & \textbf{Audio, Video, EEG, ECG}
    & \textbf{Chinese}
    & \textbf{In the community} \\
    
\bottomrule

\end{tabular}
\label{tab:dataset-review}
\end{table*}

To address these limitations, we introduce MECO, a \textbf{M}ultimodal dataset for \textbf{E}motion and \textbf{C}ognitive understanding in \textbf{O}lder adults. Collected in community-based
settings under standardized emotion elicitation protocols, MECO captures behavioral and physiological responses in ecologically valid conditions. The dataset synchronizes records of around 38 hours of multimodal signals from 42 older participants, comprising 27 healthy controls (HC) and 15 individuals with MCI. The data covers video, audio, EEG, and ECG modalities, yielding a total of 30,592 samples. Motivated by interactions between emotional responses and cognitive performance, MECO provides not only comprehensive annotations of emotional states, including self-assessed valence, arousal, and six basic emotions, but also cognitive scores based on the Mini-Mental State Examination (MMSE). Therefore, MECO supports a range of downstream tasks, including emotion–cognition modeling in aging populations, robust emotion recognition, and emotion-assisted cognitive impairment screening. Our contributions are summarized as follows:
\begin{itemize}
\item To the best of our knowledge, we present the first multimodal dataset for older adults that jointly models emotion and cognitive states. It integrates diverse behavioral and physiological modalities, addressing the lack of resources capturing emotion and cognition within aging populations.
\item We establish baseline benchmarks for emotional and cognitive prediction, demonstrating the feasibility of multimodal modeling and providing standardized evaluation protocols.
\item MECO provides a valuable resource for advancing affective computing in elderly populations, enabling the study of emotion–cognition interactions and supporting robust emotion recognition models under cognitive decline.
\end{itemize}

\section{Related Work}
\textbf{Emotion Recognition} Emotion Recognition (ER) infers human emotions from behavioral and physiological signals \cite{Koelstra2012,Soleymani2012,WeiLongZheng2015}. Multimodal approaches that integrate complementary cues outperform unimodal methods by capturing information absent in individual modalities \cite{Zhang2024,Soleymani2012}. However, most studies focus on young or middle-aged populations, leaving older adults and individuals with MCI underrepresented. Age-related changes in physiological responses and behavior introduce ER challenges, such as altered EEG signatures and diminished facial expressivity \cite{Poria2017}. Multimodal fusion in these demographics is further hindered by increased signal noise and high inter-subject variability. Consequently, multimodal ER approaches for older adults are urgently needed to enable accurate emotion prediction and support downstream applications, including mental health monitoring and cognitive care.

\textbf{Multimodal Emotion Dataset}
Table~\ref{tab:dataset-review} summarizes representative multimodal emotion datasets. Despite their contributions, several limitations remain for geriatric and emotion-cognition research.
First, concerning age distribution, most datasets (\textit{e.g.,} DEAP \cite{Koelstra2012}, SEED-IV \cite{WeiLongZheng2015}, and MAHNOB-HCI \cite{Soleymani2012}) predominantly feature young adults. While ElderReact \cite{Ma20191} targets older populations, it focuses solely on cognitively healthy individuals and lacks the physiological modalities necessary to investigate internal affective mechanisms. 
Second, existing datasets typically provide isolated emotional annotations. Although some datasets offer multi-label annotations, such as EMOTyDA \cite{Saha2020} and MINE \cite{Yang2025} (emotion and intention), cognitive assessments are generally absent, limiting emotion-cognition interaction studies.
Third, a trade-off persists between ecological validity and signal richness. In-the-wild datasets (\textit{e.g.,} DFEW \cite{Jiang2020}, MINE \cite{Yang2025}) lack physiological data, whereas lab-based ones (\textit{e.g.,} DREAMER \cite{Katsigiannis2018}, DEAP \cite{Koelstra2012}) provide high-quality recordings but may not reflect real-world responses.
To bridge these gaps, the MECO dataset provides synchronized behavioral and physiological signals with joint emotion and cognitive annotations for older adults, including those with MCI.

\section{MECO Dataset}
As shown in Fig.~\ref{fig-overview}, MECO dataset consists of approximately 38 hours of multimodal recordings from 42 participants, resulting in 30,592 samples, with details on acquisition, annotation, statistics, and ethical considerations. We introduce MECO to support studies of the interplay between emotion and cognition in older adults. 
\subsection{Data Acquisition}
\textbf{Emotion-Induction Videos}
Emotion elicitation was performed using the Emotion-Inducing Video Dataset \cite{Liang2025}, designed for Chinese older adults. The stimuli have been validated through subjective and physiological measures, ensuring high inter-rater reliability and effective elicitation. The dataset includes six discrete emotions, each with three distinct events to guarantee stimulus diversity (duration distributions are detailed in Fig.~\ref{fig-overview} (a)). These age-appropriate, culturally relevant, and safety-screened stimuli provide a standardized and ecologically valid foundation for affective data acquisition.

\textbf{Equipment and Setup}
Data acquisition was conducted using a portable tablet-based platform \cite{Zhao2025TNSRE} to synchronously capture behavioral and physiological signals (see Fig.~\ref{fig-overview} (b-ii)). Video ($1920 \times 1080$ resolution at 30 fps) and audio streams were temporally aligned with physiological recordings. Single-channel ECG and dual-channel prefrontal EEG (Fp1, Fp2) were recorded at 250 Hz. This unobtrusive wearable setup ensures high-fidelity acquisition while minimizing the physical burden placed upon older adults. All multimodal recordings were performed in a semi-controlled, naturalistic indoor environment, ensuring a quiet and familiar setting for participants.

\begin{figure}[!t]
\centering
\includegraphics[width=3.3in]{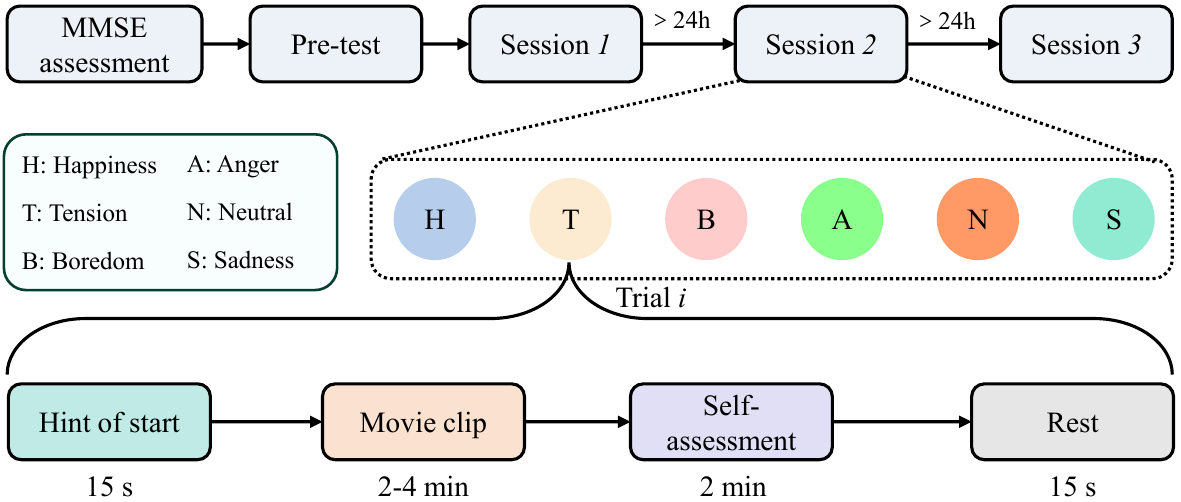}
\caption{Overview of the experimental protocol.}
\Description{A schematic of the experimental protocol showing the sequential stages of the data collection process, including participant preparation, presentation of emotion-eliciting stimuli, synchronized recording of multimodal signals, and post-stimulus self-assessment and cognitive evaluation.}
\label{fig-exp-protocol}
\end{figure}

\textbf{Experimental Protocol}
As illustrated in Fig.~\ref{fig-exp-protocol}, the experimental protocol consisted of several sequential phases, began with the MMSE assessment \cite{ArevaloRodriguez2021}, followed by a pre-test phase. Participants then completed three sequential sessions, each separated by intervals exceeding 24 hours to mitigate carryover effects and emotional fatigue. Each session consisted of six trials, corresponding to the induction of six emotions  (\textit{e.g.,} \textit{anger, boredom, happiness, neutral, sadness, tension}). In each trial, participants first received a 15-second prompt and then viewed a 2–4 minute emotion-inducing video, during which video, EEG, and ECG signals were synchronously recorded (see Fig.~\ref{fig-overview} (c)). Subsequently, participants completed a 2-minute self-assessment, with audio recorded alongside video, EEG, and ECG signals. Each trial concluded with a 15-second rest period before proceeding to the next trial.

\textbf{Participants}
Participants were elderly native Chinese speakers with underlying health conditions. To ensure sample consistency and reduce confounding effects, inclusion criteria required participants to be aged 50 years or older, capable of independent daily living, and able to provide informed consent. Exclusion criteria included severe neurological or psychiatric disorders (\textit{e.g.,} cerebrovascular diseases, schizophrenia, or severe depression), major systemic illnesses (\textit{e.g.,} hepatic or renal insufficiency), and communication impairments that could hinder compliance with the protocol. Initially, 102 participants were recruited. After accounting for technical anomalies and incomplete participation, the final MECO dataset comprises 42 subjects (27 HC and 15 MCI) who completed all three recording sessions. Fig~\ref{fig-overview} (d) summarizes the demographic characteristics and MMSE assessment results.

\subsection{Data Annotation}
To ensure reliable and reproducible labels, both cognitive status and emotional states were systematically annotated.

\textbf{Cognitive Annotation} 
Participants were dichotomized into MCI (MMSE score $\leq$ 26) and HC (MMSE score > 26). This threshold serves as a practical criterion for cognitive stratification, consistent with prior studies employing MMSE as a screening tool \cite{Folstein1975,Liang2025}.

\textbf{Emotion Annotation}
Emotional responses were collected via a hybrid  combining categorical-dimensional scheme (Fig.~\ref{fig-overview} b-iii). Immediately post-stimulus, participants self-reported six discrete categories alongside 9-point Likert ratings (1--9) for valence (negative to positive) and arousal (calm to excited). To mitigate ambiguity from perceptual and physiological overlap in low-arousal states, boredom was merged into the neutral category \cite{Liang2025}.

\textbf{Annotation Reliability}
To quantify consistency, the intraclass correlation coefficient (ICC) was computed via a two-way random-effects model \cite{Shrout1979}. The high average-measure reliability (ICC(2,$k$)) for valence (0.9855--0.9901) and arousal (0.8962--0.9691) confirms strong consistency in aggregated annotations (Table~\ref{tab:icc_results}). Conversely, the lower single-measure reliability (ICC(2,1)) reflects inter-subject variability in emotional perception, particularly among older adults.

\begin{table}[!t]
\centering
\footnotesize
\caption{Continuous emotion annotation ICCs across three sessions. Both single-rater consistency ICC(2,1) and average-rater reliability ICC(2,$k$) are highly significant ($p < 0.001$).}
\begin{tabular}{llcccc}
\toprule
\multirow{2}{*}{\textbf{Session}} & \multirow{2}{*}{\textbf{Dimension}} & \multicolumn{2}{c}{\textbf{Single-Rater ICC(2,1)}} & \multicolumn{2}{c}{\textbf{Average-Rater ICC(2,$k$)}} \\
\cmidrule(lr){3-4} \cmidrule(lr){5-6}
& & \textbf{ICC} & \textbf{95\% CI} & \textbf{ICC} & \textbf{95\% CI} \\
\midrule
\multirow{2}{*}{Session 1} 
& Valence & 0.7038 & [0.47, 0.94] & 0.9901 & [0.97, 1.00] \\
& Arousal & 0.4272 & [0.21, 0.82] & 0.9691 & [0.92, 0.99] \\
\midrule
\multirow{2}{*}{Session 2} 
& Valence & 0.6875 & [0.45, 0.93] & 0.9893 & [0.97, 1.00] \\
& Arousal & 0.3250 & [0.15, 0.75] & 0.9529 & [0.88, 0.99] \\
\midrule
\multirow{2}{*}{Session 3} 
& Valence & 0.6182 & [0.38, 0.91] & 0.9855 & [0.96, 1.00] \\
& Arousal & 0.1705 & [0.06, 0.57] & 0.8962 & [0.74, 0.98] \\
\bottomrule
\end{tabular}
\label{tab:icc_results}
\end{table}

\subsection{Dataset Statistics}
\begin{figure}[!t]
\centering
\includegraphics[width=3.2in]{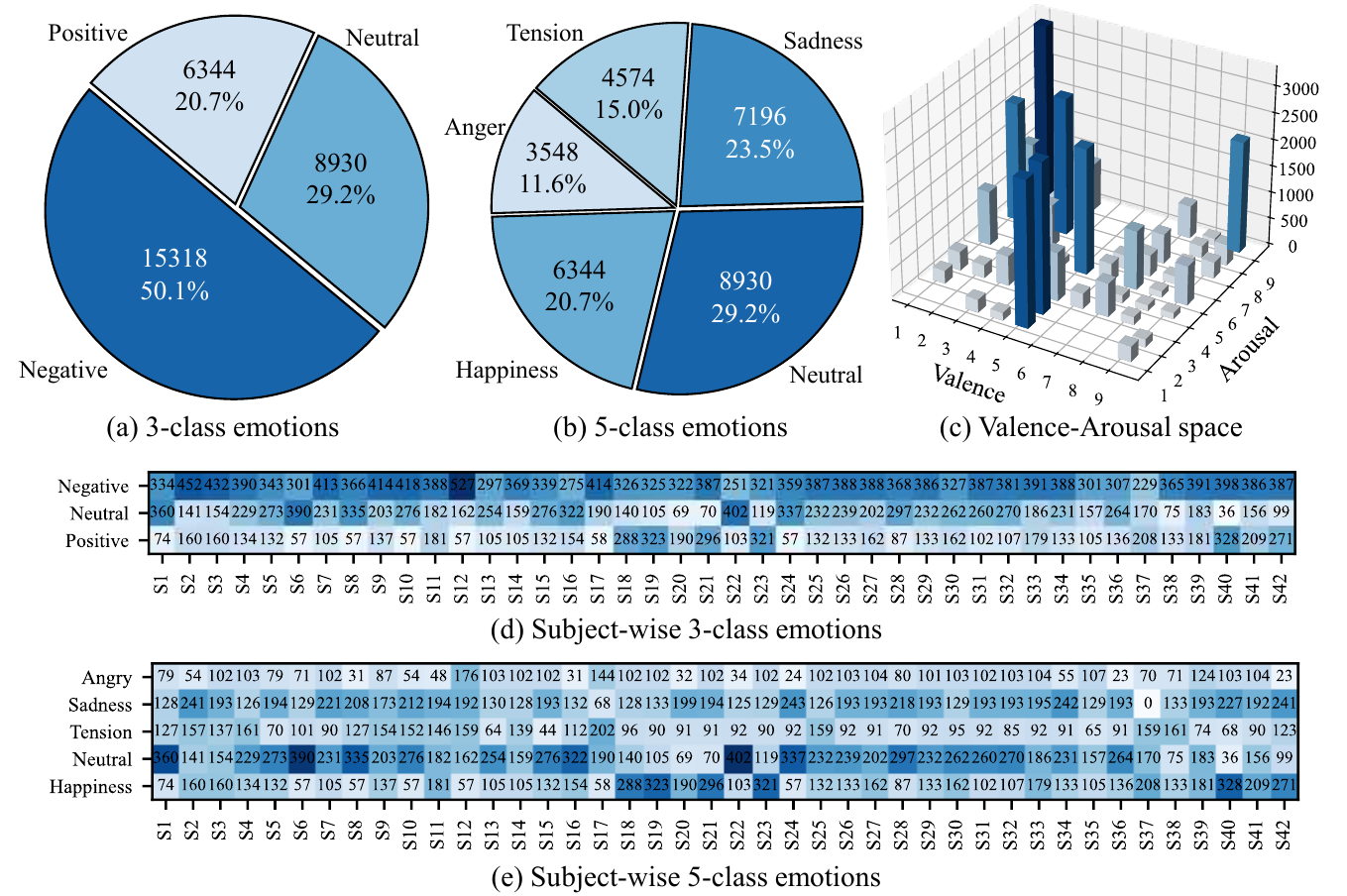}
\caption{
Distribution of MECO dataset. 
(a-c) Global-level statistics, including 3- and 5-class discrete emotion distributions and the continuous valence-arousal space. 
(d, e) Subject-level emotion distributions under 3- and 5-class settings.
}
\Description{A set of plots showing the data distribution in the MECO dataset. The first three panels present global statistics, including class distributions under different discrete emotion settings and the distribution of samples in the valence-arousal space. The last two panels illustrate subject-level variations, showing how emotion labels are distributed across individuals under the same classification schemes.}
\label{fig-distribution}
\end{figure}

Fig.~\ref{fig-distribution} details the MECO dataset statistics. Globally, the dataset exhibits a natural class imbalance characteristic of authentic emotion elicitation. Negative emotions (\textit{e.g.,} sadness, tension) dominate both the 3-class (Fig.~\ref{fig-distribution} (a)) and 5-class (Fig.~\ref{fig-distribution} (b)) settings, posing a challenging yet practical scenario for training robust emotion recognition models. Furthermore, the continuous valence-arousal distribution (Fig.~\ref{fig-distribution} (c)) is consistent with the discrete labels, showing dense clusters aligned with dominant affective states. Subject-level heatmaps (Fig.~\ref{fig-distribution} (d, e)) reveal substantial inter-subject variability. Although overall biased toward negative states, individual responses vary markedly, highlighting the dataset as a benchmark for evaluating model generalization and personalized prediction.

\subsection{Ethics Review and License}
This study was approved by the Institutional Ethics Committee (No. KY2022784), and all participants gave written informed consent. All privacy-sensitive information is protected, and anonymity is strictly guaranteed. The dataset is released under CC BY 4.0 license, permitting academic and commercial reuse with attribution.

\section{Baseline Experiments}
\label{sec:benchmark}
\subsection{Task Definition}
We formulate predictive tasks across emotion and cognition dimensions. Let the dataset be denoted as $\mathcal{\bm{D}} = \{(\bm{x}_i, e_i, v_i, a_i, c_i, m_i)\}_{i=1}^{N}$, where $\bm{x}_i$ represents the multimodal recording of the $i$-th sample among $N$ samples. For emotion prediction, $e_i \in \{1, 2, \dots, C\}$ denotes the discrete emotion category, which $v_i, a_i \in [1, 9]$ denote continuous valence and arousal scores. For cognitive prediction, $c_i \in \{0, 1\}$ indicates binary cognitive impairment status, and $m_i$ denotes the continuous MMSE score. 

Based on $\mathcal{\bm{D}}$ and Fig.~\ref{fig-overview} (e), we define five emotion-related tasks: T1 (SR), stimulus-induced emotion recognition using intended stimulus labels; T2 (SA), 3-class sentiment analysis; T3 (ER), 5-class emotion recognition; T4 (VR), valence regression; and T5 (AR), arousal regression. In addition, two cognition-related tasks are defined: T6 (CR), binary cognitive impairment recognition; and T7 (MR), MMSE score regression. Notably, the baseline tasks focus on spontaneous responses elicited during stimulus viewing, utilizing video, EEG, and ECG modalities.

\subsection{Feature Extraction}
\textbf{Data Preprocessing} Before extracting features, we apply preprocessing steps to enhance signal quality. For video data, we uniformly sample 16 frames from each clip \cite{Tran2015,BagherZadeh2018,bertasius2021space,Zhang2024CVPR}. Facial regions are then detected and aligned using OpenFace \cite{Baltrusaitis2016}, and the cropped face images are resized to $224 \times 224$. For EEG signals, a 0.5-50 Hz band-pass filter is applied  \cite{WeiLongZheng2015}. For ECG signals, baseline wander is removed and amplitude bias is mitigated using median filtering \cite{Hsu2020}, followed by 0.5-45 Hz band-pass filtering and Z-score normalization. To enable multimodal alignment and fusion, all modalities are segmented into non-overlapping 4-second sliding windows \cite{Zheng2019}.

\textbf{Video Modality}
1) Action Units (AU): We leverage the OpenFace \cite{Baltrusaitis2016} to extract 35 AUs capturing facial muscle dynamics \cite{Zhao2025}. For each segment, the mean, standard deviation, and delta mean are computed, yielding a 105-D feature vector.
2) Head Pose (HP): We extract the six degrees of freedom of head pose \cite{Valstar2016,Sen2023} and compute the same statistical and temporal descriptors, resulting in an 18-D feature vector.
3) Eye Gaze (EG): Two gaze angles are extracted and processed with the same temporal descriptors, producing a 6-D feature vector.
4) Deep Features (DF): We extract 512-D frame-level features using a ResNet-50 pretrained on the wild-FER dataset \cite{Ryumina2022} to capture high-level spatial representations.

\textbf{EEG Modality}
1) Differential Entropy (DE): DE \cite{LiChenShi2013} is extracted across five frequency bands to characterize logarithmic energy distribution, yielding a 10-D representation of band-specific patterns.
2) Power Spectral Density (PSD): PSD \cite{Jenke2014,WeiLongZheng2015} is computed over the same bands, producing a 10-D vector reflecting spectral power linked to emotional arousal. 3) Higuchi Fractal Dimension (HFD): Given the chaotic, non-stationary EEG, we compute HFD \cite{Higuchi1988} to capture transient cognitive and morphological variations, producing a 2-D vector.
4) Sample Entropy (SE): SE \cite{Richman2000} is computed to quantify temporal irregularity, yielding a 2-D vector.

\textbf{ECG Modality}
1) Time Domain (TD): Five standard heart rate variability metrics (Mean RR, SDNN, RMSSD, NN50, and pNN50) are derived from R–R intervals, forming a 5-D vector reflecting autonomic balance and physiological correlates of emotional arousal and valence.
2) HFD: We compute HFD \cite{Higuchi1988} to quantify cardiovascular morphological irregularity and chaotic behavior, yielding a 1-D feature.
3) SE: SE \cite{Richman2000} is computed to assess cardiac structural regularity and temporal predictability, producing a 1-D  feature reflecting autonomic arousal.

\subsection{Implementation Details}
To establish baseline performance, temporally aligned multimodal features are concatenated, processed by a gated recurrent unit  \cite{Cho2014} to capture temporal dynamics, and subsequently fed into a multilayer perceptron for prediction.

All experiments were conducted on an NVIDIA RTX 3090 GPU. Models were trained for 100 epochs with a batch size of 32. Optimization is performed using AdamW \cite{loshchilov2018decoupled}, with cross-entropy loss for classification and mean squared error for regression. The learning rate was selected from ${10^{-1}, 10^{-2}, 10^{-3}, 3 \times 10^{-4}}$, combined with a cosine annealing scheduler \cite{loshchilov2017sgdr} and a weight decay of $10^{-3}$. To reduce overfitting, dropout \cite{Srivastava2014} was applied with rates in $\{0.1, 0.2, 0.3\}$, along with an early stopping mechanism.

\subsection{Evaluation Protocol and Metrics}
To comprehensively benchmark MECO dataset, we define two evaluation protocols to assess both personalized and generalized capabilities of the models: Subject-Dependent (SD) and Subject-Independent (SI). Under the SD protocol, a chronological split is applied within each trial to preserve the temporal dynamics of the elicited responses. Specifically, the first 60\% of segments in each trial are allocated for training, the next 20\% for validation, and the final 20\% for testing. 
Under the SI protocol, subject-wise five-fold cross-validation is adopted to evaluate model generalization across unseen participants. All subjects are partitioned into five disjoint subsets, with four (approximately 80\%) used for training and the remaining one (20\%) for testing in each fold.

For emotion classification tasks, unweighted average recall (UAR) and weighted average recall (WAR) \cite{ChumachenkoIG2024,Chen2025,Liu2025TAC}. For continuous emotion regression and MMSE score tasks, concordance correlation coefficient (CCC) and mean absolute error (MAE) \cite{Nicolaou2011,Ringeval2015, Chu2024} are used. Specifically, for MCI screening task, accuracy (ACC) and the macro F1-score (F1) are adopted to evaluate both overall correctness and sensitivity to the minority MCI class \cite{Weiner2011}.

\begin{table*}[!t]
\centering
\footnotesize
\caption{Results for the five emotion prediction tasks (T1–T5) on MECO dataset under the SD protocol. Best and second-best results are highlighted in bold and underlined, respectively. “M” denotes the modality: Video (V), EEG (E), and ECG (C).}
\label{emotion-sd-results}
\begin{tabular}{ll cccccccccc}
\toprule
\multirow{2}{*}{\textbf{M}} & \multirow{2}{*}{\textbf{Feature}}
& \multicolumn{2}{c}{\textbf{T1: SR} (\%)}
& \multicolumn{2}{c}{\textbf{T2: SA} (\%)}
& \multicolumn{2}{c}{\textbf{T3: ER} (\%)}
& \multicolumn{2}{c}{\textbf{T4: VR}}
& \multicolumn{2}{c}{\textbf{T5: AR}}\\
\cmidrule(lr){3-4} \cmidrule(lr){5-6} \cmidrule(lr){7-8} \cmidrule(lr){9-10} \cmidrule(lr){11-12}
& & UAR$_{\pm std}\uparrow$ & WAR$_{\pm std}\uparrow$
& UAR$_{\pm std}\uparrow$ & WAR$_{\pm std}\uparrow$
& UAR$_{\pm std}\uparrow$ & WAR$_{\pm std}\uparrow$
& CCC$_{\pm std}\uparrow$ & MAE$_{\pm std}\downarrow$
& CCC$_{\pm std}\uparrow$ & MAE$_{\pm std}\downarrow$\\
\midrule

\multirow{4}{*}{V} 
& AU & 43.09$_{\pm 10.48}$ & 45.60$_{\pm 10.10}$ & 59.20$_{\pm 9.52}$ & 62.27$_{\pm 9.68}$ & 48.00$_{\pm 10.75}$ & 52.03$_{\pm 10.40}$ & 0.4686$_{\pm 0.1729}$ & 1.7899$_{\pm 0.4995}$ & 0.4916$_{\pm 0.1463}$ & 1.7007$_{\pm 0.4011}$ \\
& HP & 44.09$_{\pm 11.38}$ & 45.94$_{\pm 11.02}$ & 64.03$_{\pm 11.26}$ & 69.13$_{\pm 9.23}$ & 52.54$_{\pm 11.24}$ & 58.14$_{\pm 9.68}$ & 0.4558$_{\pm 0.1900}$ & 1.7639$_{\pm 0.4977}$ & 0.5247$_{\pm 0.1592}$ & 1.6262$_{\pm 0.5067}$ \\
& EG & 30.73$_{\pm 7.00}$  & 32.87$_{\pm 7.14}$  & 45.79$_{\pm 8.34}$ & 54.42$_{\pm 10.18}$& 34.66$_{\pm 6.93}$  & 40.87$_{\pm 8.31}$  & 0.2153$_{\pm 0.1594}$ & 2.1855$_{\pm 0.6889}$ & 0.2314$_{\pm 0.1947}$ & 2.5147$_{\pm 1.4359}$ \\
& DF & 58.09$_{\pm 12.48}$ & 59.47$_{\pm 11.96}$ & 69.85$_{\pm 11.32}$ & 72.74$_{\pm 8.44}$ & 63.64$_{\pm 11.52}$ & 65.80$_{\pm 10.08}$ & 0.5972$_{\pm 0.1544}$ & \textbf{1.4106}$_{\pm 0.3949}$ & 0.6347$_{\pm 0.1339}$ & \underline{1.3270}$_{\pm 0.3691}$ \\
\midrule

\multirow{4}{*}{E} 
& DE & 30.54$_{\pm 7.55}$  & 33.38$_{\pm 8.05}$  & 45.85$_{\pm 7.01}$  & 53.44$_{\pm 7.38}$ & 35.10$_{\pm 6.95}$  & 41.41$_{\pm 8.12}$  & 0.2789$_{\pm 0.1548}$ & 2.0873$_{\pm 0.5138}$ & 0.2964$_{\pm 0.1718}$ & 2.0340$_{\pm 0.5656}$ \\
& PSD & 28.45$_{\pm 6.03}$ & 31.68$_{\pm 6.47}$  & 44.54$_{\pm 6.51}$  & 51.08$_{\pm 9.84}$ & 32.97$_{\pm 7.27}$  & 38.60$_{\pm 8.14}$  & 0.2503$_{\pm 0.1218}$ & 2.1433$_{\pm 0.5047}$ & 0.2637$_{\pm 0.1725}$ & 2.0628$_{\pm 0.5686}$ \\
& HFD & 23.19$_{\pm 4.11}$ & 25.25$_{\pm 5.03}$  & 40.94$_{\pm 6.40}$  & 46.26$_{\pm 10.35}$& 28.83$_{\pm 6.14}$  & 34.35$_{\pm 7.03}$  & 0.1522$_{\pm 0.1360}$ & 2.3880$_{\pm 0.7002}$ & 0.1767$_{\pm 0.1550}$ & 2.2506$_{\pm 0.7885}$ \\
& SE & 23.13$_{\pm 4.65}$ & 25.11$_{\pm 6.15}$ & 39.00$_{\pm 6.17}$ & 46.69$_{\pm 7.25}$ & 26.47$_{\pm 5.44}$  & 31.25$_{\pm 5.83}$  & 0.1376$_{\pm 0.1326}$ & 2.3028$_{\pm 0.5321}$ & 0.1675$_{\pm 0.1674}$ & 2.2104$_{\pm 0.6316}$ \\
\midrule

\multirow{3}{*}{C} 
& TD & 19.05$_{\pm 3.08}$  & 22.21$_{\pm 4.89}$  & 35.01$_{\pm 3.59}$  & 43.36$_{\pm 10.86}$& 23.12$_{\pm 4.31}$  & 28.33$_{\pm 11.56}$ & 0.0532$_{\pm 0.0816}$ & 2.7174$_{\pm 1.2650}$ & 0.0895$_{\pm 0.0950}$ & 2.3558$_{\pm 1.0149}$ \\
& HFD & 23.39$_{\pm 4.09}$ & 26.85$_{\pm 4.73}$  & 41.09$_{\pm 7.85}$  & 49.49$_{\pm 9.99}$ & 28.28$_{\pm 5.05}$  & 37.32$_{\pm 8.31}$  & 0.1634$_{\pm 0.1518}$ & 2.1319$_{\pm 0.4987}$ & 0.2397$_{\pm 0.1609}$ & 2.0763$_{\pm 0.6139}$ \\
& SE & 25.32$_{\pm 4.66}$ & 28.32$_{\pm 5.73}$ & 40.10$_{\pm 6.61}$ & 49.10$_{\pm 8.03}$ & 28.38$_{\pm 6.35}$  & 34.82$_{\pm 9.25}$  & 0.1809$_{\pm 0.1487}$ & 2.1795$_{\pm 0.6836}$ & 0.2287$_{\pm 0.1538}$ & 2.1366$_{\pm 0.9865}$ \\
\midrule

\multirow{2}{*}{VE} 
& Top-1 & 62.10$_{\pm 10.93}$ & \textbf{63.43}$_{\pm 10.35}$ & 70.48$_{\pm 10.99}$ & \underline{73.27}$_{\pm 8.23}$ & \textbf{65.83}$_{\pm 10.03}$ & \textbf{67.89}$_{\pm 9.03}$ & \underline{0.5980}$_{\pm 0.1558}$ & 1.4199$_{\pm 0.4191}$ & \textbf{0.6424}$_{\pm 0.1261}$ & 1.3433$_{\pm 0.3924}$ \\
& Top-2 & 51.94$_{\pm 11.36}$ & 53.83$_{\pm 11.04}$ & 61.13$_{\pm 11.03}$ & 66.30$_{\pm 8.90}$ & 54.59$_{\pm 10.63}$ & 59.54$_{\pm 9.00}$ & 0.4503$_{\pm 0.1738}$ & 1.8346$_{\pm 0.4796}$ & 0.5192$_{\pm 0.1418}$ & 1.6695$_{\pm 0.4593}$ \\
\midrule

\multirow{2}{*}{VC} 
& Top-1 & \textbf{62.20}$_{\pm 12.39}$ & 63.37$_{\pm 11.56}$ & \textbf{71.76}$_{\pm 10.98}$ & \textbf{73.82}$_{\pm 8.53}$ & 64.96$_{\pm 12.43}$ & \underline{67.47}$_{\pm 10.04}$ & \textbf{0.6041}$_{\pm 0.1590}$ & \underline{1.4111}$_{\pm 0.4195}$ & \underline{0.6423}$_{\pm 0.1351}$ & \textbf{1.3218}$_{\pm 0.3626}$ \\
& Top-2 & 51.20$_{\pm 11.23}$ & 52.90$_{\pm 10.74}$ & 63.03$_{\pm 10.68}$ & 67.84$_{\pm 9.52}$ & 54.80$_{\pm 12.65}$ & 59.85$_{\pm 10.81}$ & 0.4768$_{\pm 0.1981}$ & 1.7345$_{\pm 0.5187}$ & 0.5120$_{\pm 0.1804}$ & 1.7266$_{\pm 0.7182}$ \\
\midrule

\multirow{2}{*}{EC} 
& Top-1 & 34.40$_{\pm 8.22}$ & 37.13$_{\pm 8.01}$ & 50.08$_{\pm 7.07}$ & 56.91$_{\pm 8.68}$ & 39.13$_{\pm 9.56}$ & 45.58$_{\pm 9.14}$ & 0.3004$_{\pm 0.1470}$ & 2.1469$_{\pm 0.6492}$ & 0.2796$_{\pm 0.1940}$ & 2.2094$_{\pm 0.8099}$ \\
& Top-2 & 32.72$_{\pm 7.34}$ & 35.67$_{\pm 7.43}$ & 48.03$_{\pm 8.79}$ & 54.12$_{\pm 8.56}$ & 36.15$_{\pm 7.97}$ & 42.13$_{\pm 7.96}$ & 0.2673$_{\pm 0.1477}$ & 2.2401$_{\pm 0.6858}$ & 0.2704$_{\pm 0.1853}$ & 2.2250$_{\pm 0.8046}$ \\
\midrule

\multirow{2}{*}{VEC} 
& Top-1 & \underline{62.13}$_{\pm 11.19}$ & \underline{63.40}$_{\pm 10.61}$ & \underline{71.22}$_{\pm 10.92}$ & 73.25$_{\pm 8.18}$ & \underline{65.09}$_{\pm 10.37}$ & 67.17$_{\pm 9.17}$ & 0.5974$_{\pm 0.1558}$ & 1.4246$_{\pm 0.3989}$ & 0.6324$_{\pm 0.1328}$ & 1.3892$_{\pm 0.4047}$ \\
& Top-2 & 61.48$_{\pm 11.96}$ & 62.71$_{\pm 11.31}$ & 65.96$_{\pm 10.76}$ & 68.10$_{\pm 9.64}$ & 54.87$_{\pm 11.65}$ & 58.91$_{\pm 11.38}$ & 0.4694$_{\pm 0.1776}$ & 1.8320$_{\pm 0.5492}$ & 0.5081$_{\pm 0.1496}$ & 1.7054$_{\pm 0.4570}$ \\
\bottomrule
\end{tabular}


\end{table*}

\begin{table*}[t]
\centering
\footnotesize
\caption{Results for the two emotion (T2–T3) and cognitive (T6–T7) prediction tasks on MECO dataset under the SI protocol.}
\label{emotion-si-results}
\begin{tabular}{l ccccccccc}
\toprule
\multirow{2}{*}{\textbf{M}} & \multirow{2}{*}{\textbf{Feature}}
& \multicolumn{2}{c}{\textbf{T2: SA} (\%)}
& \multicolumn{2}{c}{\textbf{T3: ER} (\%)}
& \multicolumn{2}{c}{\textbf{T6: CR} (\%)}
& \multicolumn{2}{c}{\textbf{T7: MR}}\\
\cmidrule(lr){3-4} \cmidrule(lr){5-6} \cmidrule(lr){7-8} \cmidrule(lr){9-10} 
& & UAR$_{\pm std}\uparrow$ & WAR$_{\pm std}\uparrow$
& UAR$_{\pm std}\uparrow$ & WAR$_{\pm std}\uparrow$
& ACC$_{\pm std}\uparrow$ & F1$_{\pm std}\uparrow$ 
& CCC$_{\pm std}\uparrow$ & MAE$_{\pm std}\downarrow$\\
\midrule

\multirow{4}{*}{V} 
& AU & \underline{41.30}$_{\pm 2.32}$ & \textbf{49.64}$_{\pm 1.90}$ & \textbf{25.90}$_{\pm 0.76}$ & 29.60$_{\pm 2.89}$ & 60.84$_{\pm 2.44}$ & 54.28$_{\pm 3.64}$ & 0.0981$_{\pm 0.0838}$ & \underline{3.2101}$_{\pm 0.5792}$ \\
& HP & 36.66$_{\pm 1.30}$ & 47.74$_{\pm 2.55}$ & 22.67$_{\pm 1.25}$ & 28.44$_{\pm 3.32}$ & 60.51$_{\pm 4.73}$ & 51.23$_{\pm 3.89}$ & 0.0762$_{\pm 0.0357}$ & 3.4828$_{\pm 0.5989}$ \\
& EG & 35.34$_{\pm 0.37}$ & 49.34$_{\pm 1.12}$ & 22.37$_{\pm 1.20}$ & \textbf{30.46}$_{\pm 4.65}$ & 62.46$_{\pm 2.99}$ & 45.55$_{\pm 4.00}$ & 0.0228$_{\pm 0.0280}$ & 3.6050$_{\pm 0.1252}$ \\
& DF & 40.21$_{\pm 1.09}$ & 45.49$_{\pm 1.12}$ & 24.56$_{\pm 1.04}$ & 28.07$_{\pm 1.61}$ & 57.89$_{\pm 12.55}$ & 52.57$_{\pm 11.39}$ & 0.2286$_{\pm 0.1805}$ & 3.4754$_{\pm 0.7802}$ \\
\midrule

\multirow{4}{*}{E} 
& DE & 34.19$_{\pm 0.25}$ & 47.90$_{\pm 2.21}$ & 21.11$_{\pm 0.35}$ & 26.53$_{\pm 1.42}$ & 56.72$_{\pm 3.50}$ & 52.72$_{\pm 3.17}$ & 0.1401$_{\pm 0.0865}$ & 4.2936$_{\pm 2.0246}$ \\
& PSD & 35.48$_{\pm 0.71}$ & 45.89$_{\pm 2.33}$ & 21.97$_{\pm 0.73}$ & 25.54$_{\pm 2.83}$ & 56.73$_{\pm 4.20}$ & 53.16$_{\pm 2.86}$ & 0.1439$_{\pm 0.0731}$ & 3.9832$_{\pm 1.5692}$ \\
& HFD & 35.64$_{\pm 1.16}$ & 40.92$_{\pm 5.65}$ & 22.23$_{\pm 0.44}$ & 27.79$_{\pm 3.69}$ & 55.59$_{\pm 6.89}$ & 44.74$_{\pm 3.78}$ & 0.1356$_{\pm 0.0841}$ & 5.0774$_{\pm 1.2861}$ \\
& SE & 33.70$_{\pm 0.31}$ & 48.42$_{\pm 2.21}$ & 21.42$_{\pm 0.30}$ & 28.80$_{\pm 2.68}$ & 58.95$_{\pm 4.84}$ & 42.60$_{\pm 1.16}$ & 0.0991$_{\pm 0.0516}$ & 3.4277$_{\pm 0.8076}$ \\

\midrule

\multirow{3}{*}{C} 
& TD & 33.35$_{\pm 0.03}$ & \underline{49.59}$_{\pm 2.00}$ & 20.66$_{\pm 0.92}$ & 28.00$_{\pm 4.37}$ & 62.28$_{\pm 4.15}$ & 42.24$_{\pm 6.64}$ & 0.0105$_{\pm 0.0136}$ & 7.7322$_{\pm 6.5972}$ \\
& HFD & 37.08$_{\pm 1.80}$ & 35.20$_{\pm 10.81}$ & 22.45$_{\pm 0.33}$ & 24.90$_{\pm 3.93}$ & \underline{62.79}$_{\pm 2.76}$ & 44.04$_{\pm 5.32}$ & 0.2086$_{\pm 0.1065}$ & 5.3417$_{\pm 2.0126}$ \\
& SE & 34.28$_{\pm 1.26}$ & 49.22$_{\pm 1.79}$ & 21.58$_{\pm 0.95}$ & 29.66$_{\pm 4.06}$ & \textbf{63.71}$_{\pm 2.81}$ & 41.76$_{\pm 3.65}$ & 0.1221$_{\pm 0.1029}$ & 4.4568$_{\pm 0.4578}$ \\
\midrule

\multirow{2}{*}{VE} 
& Top-1 & \textbf{41.55}$_{\pm 2.39}$ & 48.35$_{\pm 2.09}$ & \underline{25.58}$_{\pm 0.81}$ & 30.26$_{\pm 2.49}$ & 59.23$_{\pm 1.75}$ & \textbf{54.58}$_{\pm 2.63}$ & 0.2626$_{\pm 0.2837}$ & \textbf{3.1895}$_{\pm 0.9175}$ \\
& Top-2 & 40.00$_{\pm 1.08}$ & 45.45$_{\pm 1.61}$ & 23.98$_{\pm 1.52}$ & 26.62$_{\pm 2.41}$ & 56.17$_{\pm 3.33}$ & 53.15$_{\pm 3.55}$ & 0.1015$_{\pm 0.1541}$ & 3.3886$_{\pm 0.7263}$ \\

\midrule

\multirow{2}{*}{VC} 
& Top-1 & 41.24$_{\pm 1.84}$ & 48.47$_{\pm 2.86}$ & 25.37$_{\pm 1.39}$ & \underline{30.36}$_{\pm 1.40}$ & 60.29$_{\pm 2.60}$ & 53.88$_{\pm 4.45}$ & \textbf{0.2812}$_{\pm 0.2800}$ & 3.3263$_{\pm 0.9776}$ \\
& Top-2 & 39.76$_{\pm 1.08}$ & 45.85$_{\pm 1.86}$ & 24.16$_{\pm 0.90}$ & 27.15$_{\pm 1.11}$ & 59.61$_{\pm 6.54}$ & 50.55$_{\pm 5.04}$ & 0.1424$_{\pm 0.1433}$ & 3.3833$_{\pm 0.5293}$ \\

\midrule

\multirow{2}{*}{EC} 
& Top-1 & 35.31$_{\pm 1.20}$ & 47.57$_{\pm 2.49}$ & 22.65$_{\pm 0.84}$ & 26.71$_{\pm 2.19}$ & 55.09$_{\pm 4.49}$ & 51.00$_{\pm 3.52}$ & 0.2371$_{\pm 0.1100}$ & 3.8719$_{\pm 1.2773}$\\
& Top-2 & 35.11$_{\pm 0.72}$ & 41.61$_{\pm 7.26}$ & 22.64$_{\pm 0.56}$ & 26.37$_{\pm 2.18}$ & 56.98$_{\pm 2.78}$ & 53.07$_{\pm 2.26}$ & 0.1655$_{\pm 0.0726}$ & 4.1042$_{\pm 1.4751}$ \\

\midrule

\multirow{2}{*}{VEC} 
& Top-1 & 40.16$_{\pm 1.76}$ & 47.64$_{\pm 2.00}$ & 25.13$_{\pm 0.17}$ & 28.79$_{\pm 1.57}$ & 59.24$_{\pm 3.08}$ & \underline{54.53}$_{\pm 2.98}$ & \underline{0.2795}$_{\pm 0.2602}$ & 3.3568$_{\pm 1.0216}$ \\
& Top-2 & 39.75$_{\pm 1.89}$ & 45.72$_{\pm 3.15}$ & 24.21$_{\pm 1.31}$ & 27.77$_{\pm 1.38}$ & 55.59$_{\pm 3.38}$ & 50.30$_{\pm 4.93}$ & 0.1387$_{\pm 0.1755}$ & 3.6860$_{\pm 0.7725}$ \\

\bottomrule
\end{tabular}



\end{table*}

\section{Results and Analysis}
\subsection{Emotion Prediction}
Table~\ref{emotion-sd-results} reports the performance under the SD protocol across five emotion tasks, where the two best-performing features from each modality are selected for fusion. 
1) For unimodal evaluation, video modality yields the most competitive overall performance. DF consistently outperforms handcrafted features, indicating that high-dimensional representations are essential for capturing subtle facial dynamics. For physiological signals, performance is modality-dependent. 
For EEG modality, DE and PSD achieve the best results, suggesting that emotional variations are more effectively captured by localized frequency-band energy than by global non-linear complexity. For ECG modality, non-linear features consistently outperform TD features, as linear statistics tend to smooth rapid autonomic fluctuations associated with short-term stimuli.
2) For multimodal evaluation, the results demonstrate clear cross-modal complementarity. Bimodal combinations (\textit{e.g.,} VE and VC) consistently surpass all unimodal baselines, confirming that integrating facial behaviors with physiological signals enhances prediction. However, trimodal fusion does not yield further improvements and may even degrade performance, suggesting that direct concatenation introduces redundancy and cross-modal interference.

Table~\ref{emotion-si-results} presents results under the SI protocol. Unimodal performance, particularly from the video modality, provides a strong baseline. Multimodal results are comparable to or slightly lower than the best unimodal outcomes. This degradation primarily stems from the strict cross-subject generalization setting, where substantial inter-subject variability in facial expressions and physiological responses exists. When evaluated on unseen subjects, feature concatenation fails to capture cross-modal representations and instead amplifies heterogeneous subject-specific noise.

\subsection{Cognitive Prediction}
Table~\ref{emotion-si-results} reports the cognitive prediction results. 1) Similar to emotion tasks, video modality provides a strong unimodal baseline. Although ECG features reach the highest ACC for T6 (63.71\%), its lower F1 indicates a bias toward the majority class, highlighting video representations as more stable indicators under the SI protocol. EEG features provide moderate but consistent contributions across both tasks. 
2) Multimodal fusion demonstrates promising potential, particularly for continuous cognitive assessment. For T7, multimodal integration yields the best overall performance, with bimodal combinations such as VC and VE achieving highest CCC and lowest MAE, respectively. This suggests that physiological signals provide complementary information for fine-grained cognitive tracking. For T6, multimodal configurations perform comparably to the strong video baseline. Rather than indicating modality limitations, this plateau under the strict SI protocol reflects the profound individual heterogeneity inherent in the physiological responses among older adults. Simple feature concatenation is insufficient to disentangle these complex individual differences, highlighting the need for domain-adaptive or context-aware fusion strategies to better exploit cross-modal synergies.

\section{Conclusion}
In this work, we introduced MECO, the first multimodal dataset dedicated to emotion and cognitive understanding in older adults, which integrates behavioral and physiological signals with unified annotations for affective states and cognitive assessment. We established baseline benchmarks for both emotion and cognitive prediction under unimodal and multimodal settings, providing a standardized reference for reproducible evaluation. 
However, the current study has certain limitations. Specifically, the analysis is restricted to stimulus-elicited data, leaving the audio modality unanalyzed, and the dataset is limited to 42 subjects. We plan to integrate audio data and recruit more subjects for further analysis. 
Furthermore, MECO offers a rich multimodal foundation to pre-train robust emotion recognition models for geriatric populations, ultimately advancing cognitively aware intelligent systems.
\newpage






\bibliographystyle{ACM-Reference-Format}
\bibliography{reference}










\end{document}